\documentclass[sigconf]{acmart}
\renewcommand\footnotetextcopyrightpermission[1]{} 

\usepackage{enumitem}

\usepackage[nameinlink, noabbrev, capitalize]{cleveref}
\usepackage{graphicx} 
\usepackage{orcidlink} 
\usepackage{xcolor}
\usepackage{breakurl}
\usepackage{natbib}
\usepackage{tabularx}
\usepackage{ragged2e}
\newcolumntype{L}[1]{>{\RaggedRight\arraybackslash}p{#1}}
\newcolumntype{Y}{>{\RaggedRight\arraybackslash}X}
\usepackage{booktabs}

\PassOptionsToPackage{hyphens}{url}\usepackage{hyperref}

\copyrightyear{2026}
\acmYear{2026}
\setcopyright{cc}
\setcctype{by}
\acmConference[SIGCSE TS 2026]{Proceedings of the 57th ACM Technical Symposium on Computer Science Education V.1}{February 18--21, 2026}{St. Louis, MO, USA}
\acmBooktitle{Proceedings of the 57th ACM Technical Symposium on Computer Science Education V.1 (SIGCSE TS 2026), February 18--21, 2026, St. Louis, MO, USA}
\acmDOI{XXXXXXX.XXXXXXX}
\acmISBN{978-1-4503-XXXX-X/2018/06}

\usepackage{xspace}
\newcommand{\glow}{GLOW\xspace}
\newcommand{\owlgorithm}{Owlgorithm\xspace}
\newcommand{\boilertai}{BoilerTAI\xspace}
\newcommand{\boilersketch}{BoilerSketch\xspace}
\newcommand{\codestylist}{CodeStylist\xspace}
\newcommand{\ailab}{AI-Lab\xspace}
\newcommand{\gaide}{GAIDE\xspace}
\title{Instructional Governance by Design: A Framework for AI in Computing Education}

\author{Ethan Dickey}
\affiliation{
  \institution{Purdue University}
  \city{West Lafayette}
  \state{Indiana}
  \country{USA}
  }
\email{dickeye@purdue.edu}
\orcid{0009-0007-3706-5253}

\begin{abstract}
    As generative AI permeates computing instruction, the emergent design challenge is to configure each tool's pedagogical role, authority, and accountability for the instructional work it performs. We argue for instructional governance by design: governance should be encoded in a teaching tool's interaction model, constraints, and workflow. We introduce a multidimensional framework that characterizes AI teaching tools through (1) pedagogical grounding, (2) AI instructional authority, (3) human accountability and control, (4) learner agency and cognitive engagement, (5) context specificity and boundary setting, and (6) evaluation visibility and revision. These dimensions yield governance profiles that help educators align tools with specific purposes and educational stakes. We develop the position through a comparative analysis of a portfolio of AI teaching tools across computing and first-year engineering: rubric-anchored GTA simulations, reflection-oriented code companions, staff-reviewed forum-response systems, TA-supervised diagram generators, and course-specific code-style coaches. These cases show how common instructional functions call for different combinations of rubrics, learning-theory commitments, approval gates, supervision, and course-specific constraints. We further apply the framework to selected published tools to demonstrate its use beyond a single institutional portfolio. From these cases, we identify reusable design questions for aligning governance with instructional stakes, human capacity, and intended learning processes. This position reframes responsible AI integration as a curricular and interaction-design challenge and offers a common vocabulary for tool builders, instructors, and researchers to design, compare, and evaluate AI-mediated learning environments.
\end{abstract}

\begin{document}

\ccsdesc[500]{Social and professional topics~Computing education}

\keywords{Generative AI, AI governance, Computing education, Instructional governance, Interaction design, Educational technology}

\maketitle
\pagestyle{plain} 





\section{Introduction}
\label{sec:introduction}

Generative AI (genAI) has shifted computing education from an era of scarce help to an era of abundant, immediate, and plausible help. A single large language model can explain compiler errors, generate code, critique style, summarize documentation, draft forum replies, simulate student questions, and produce examples or diagrams \cite{finnieansley2022robots,becker2023programming,denny2024era}. For students, this abundance changes the conditions under which they practice programming, debug misconceptions, ask for help, and demonstrate understanding \cite{chen2024plagiarism}. For instructors and teaching assistants, it changes the economics of feedback, the boundaries of acceptable assistance, and the forms of instructional labor that can be scaled. The central design question for computing education is therefore no longer simply whether genAI should be permitted in a course. The deeper question is how AI-mediated instructional interactions should be organized so that scale, responsiveness, and personalization support rather than displace the learning processes educators intend to cultivate.

The first wave of computing education research on genAI established the breadth of this disruption. Community syntheses have documented rapidly expanding model capabilities, educator and student attitudes, ethical concerns, policy challenges, and emergent adaptations across courses and institutions \cite{denny2024era,prather2025beyond,cambaz2024use}. Empirical work has shown that instructors are actively rethinking assignments, assessments, and course policies in response to AI code generation tools \cite{sheard2024instructor,zastudil2023perspectives,smith2024early}. Studies of novice programming contexts have shown both the potential and the instructional risk of genAI: students can receive explanations, hints, and working code at unprecedented speed, while novice metacognitive difficulties and overconfidence can make it hard for learners to judge whether generated assistance is correct, useful, or educationally appropriate \cite{denny2023conversing,hellas2023exploring,loksa2022metacognition,prather2024widening,prather2023copilot,rahe2025how}. These findings collectively point to a durable challenge for the field: genAI changes not only what students can produce, but also how they regulate, explain, evaluate, and take responsibility for their own learning activity.

The community is also moving from general-purpose chatbot use toward purpose-built AI teaching tools.
Digital teaching assistants, guarded programming-help systems, hint generators, prompt-construction exercises, personalized Parsons-puzzle generators, code explanation systems, code-style feedback systems, worked-example generators, and notebook interfaces for AI-assisted code generation demonstrate that educational value depends on interaction design, not just on the presence of a powerful model \cite{sarsa2022automatic,macneil2023experiences,denny2024desirable,liffiton2024codehelp,kazemitabaar2024codeaid,denny2024prompt}. Recent systems further show how LLM-based support can be governed through scaffolded tasks, guarded hints, contextual tutoring, interface mediation, and large-course deployment workflows \cite{hou2024codetailor,woodrow2024aiteaches,cheng2024biscuit,roest2024nextstep,bassner2024iris,zamfirescu2025bot}. These systems do more than answer questions. They decide, through their design, what counts as help, how much solution information is appropriate, when learners should remain active constructors, how much control users retain, and what traces are available to students or staff afterwards. In this sense, many key design choices in AI tools are governance choices.


We use \emph{instructional governance} to name the designed allocation of pedagogical authority, learner agency, human accountability, contextual boundaries, and evaluation visibility within an AI-mediated teaching interaction. This definition places governance at the level where computing educators and students encounter AI: the prompt, rubric, approval gate, interface, dashboard, style guide, scenario bank, feedback workflow, or course-specific constraint. Recent work on AI governance in computing education has argued for discipline-aware policy design, transparency, tiered permissions, integrity evaluation, equitable access, and clearer communication of permitted uses \cite{jamil2026governance,ali2025policies}. Human-AI interaction research has likewise emphasized that effective AI systems require calibrated expectations, visible uncertainty, meaningful control, and support for user correction \cite{amershi2019guidelines}. Instructional governance by design extends these concerns into the architecture of learning tools. It asks how an AI is pedagogically grounded, what instructional role it is permitted to occupy, who remains accountable for consequential outputs, what cognitive work learners retain, how local course context constrains the interaction, and how the system can be inspected and revised.

Our position is that responsible scale in genAI-supported computing education depends on \emph{governance fit}: coherence between the instructional function of a tool and the governance mechanisms in its design. Tools that draft official course-forum replies, simulate a confused student for GTA rehearsal, ask reflection questions about submitted code, generate diagrams, and critique code style should not be governed by the same interaction pattern. Each assigns different authority to the AI system, different responsibility to humans, and different intellectual work to learners. Constructive alignment reminds us that learning activities, assessments, and goals should be coherently organized \cite{biggs1996constructive}; technological pedagogical content knowledge similarly emphasizes that educational technologies become meaningful through their relationship to content and pedagogy \cite{mishra2006tpck}. Instructional governance applies this design logic to genAI: the model, interface, workflow, and evaluation traces must be aligned with the learning activity the tool is meant to support.

This paper makes four contributions. First, we articulate instructional governance as a design-level construct for computing education, complementing institutional policy and course-level AI-use rules. Second, we introduce a governance profile framework with six dimensions: pedagogical grounding, AI instructional authority, human accountability and control, learner agency and cognitive engagement, context specificity and boundary-setting, and evaluation visibility and revision. Third, we use design vignettes from a portfolio of AI teaching tools and selected external systems to show how different instructional functions call for different governance profiles. Fourth, we identify transferable design questions for responsible scale, arguing that educators and researchers should evaluate AI tools by the learning activity and accountability structure they create, not only by their output quality or model capability.

By locating governance in instructional artifacts and workflows, this paper reframes responsible AI integration as a problem of curriculum, pedagogy, and interaction design. The goal is not to produce a universal permission policy for genAI in computing courses. The goal is to give computing educators a vocabulary for designing, comparing, and studying AI-mediated learning environments in which authority, agency, accountability, context, and evidence are deliberately arranged. Such a vocabulary is increasingly necessary as AI teaching tools become embedded in the everyday infrastructure of programming education, engineering education, and the preparation of the instructional staff who support them.

\section{A Governance Profile Framework}
\label{sec:framework}

To make instructional governance actionable for computing educators, we define a \emph{governance profile} as an explicit account of how an AI-mediated teaching tool distributes pedagogical authority, learner agency, human accountability, contextual boundaries, and evaluation visibility across the people, artifacts, and computational systems involved in an instructional interaction. A governance profile is intended to function as a design representation: it describes what the AI system is allowed to do, what pedagogical commitments constrain its behavior, what work remains with students or staff, and how the interaction can be inspected and improved.

This framing shifts responsible AI integration from a policy-only concern to a course-design and interaction-design concern. Institutional policies can specify permissible use, privacy expectations, or academic integrity requirements, but students and instructional staff encounter AI through concrete workflows: prompts, rubrics, approval gates, interface affordances, dashboards, style guides, and feedback cycles. These artifacts determine whether an AI acts as a coach, simulator, drafter, evaluator, tutor, content generator, or object of critique. The governance profile makes those assignments visible before deployment and available for revision after use.

\Cref{tab:profile-dimensions} summarizes six dimensions of instructional governance. The dimensions are analytically separable, though they interact in practice. For example, a tool may be strongly grounded in learning theory while giving the AI system little authority over final outputs, or it may be highly exploratory while still being constrained by a local course style guide. This multidimensional structure avoids collapsing governance into a single continuum from ``strict'' to ``permissive.'' In educational settings, the more important question is whether the governance profile fits the instructional function.

\begin{table*}[t]
    \centering
    \small
    \begin{tabularx}{\textwidth}{p{0.13\textwidth}p{0.22\textwidth}p{0.31\textwidth}p{0.27 \textwidth}}
        \toprule
        \textbf{Dimension} &
        \textbf{Core design question} &
        \textbf{Typical mechanisms} &
        \textbf{Evidence of fit} \\
        \midrule
        
        \begin{tabular}[t]{@{}l@{}} Pedagogical\\grounding \end{tabular} &
        What educational goal, theory, rubric, or course standard constrains the tool? &
        Learning objectives; instructor-authored rubrics; Bloom- or SRL-aligned prompts; course-specific style guides; constructivist or TPACK-informed content workflows &
        Alignment between outputs and learning goals; rubric consistency; instructor judgments of pedagogical relevance \\
        
        \addlinespace[0.25em]
        
        \begin{tabular}[t]{@{}l@{}} AI instructional\\authority \end{tabular} &
        What instructional role is the AI system permitted to occupy? &
        AI as simulator, question generator, drafter, diagram generator, code-style coach, evaluator, or content development partner &
        Role clarity; appropriateness of output stakes; frequency and severity of problematic outputs; student and staff understanding of the AI's role \\
        
        \addlinespace[0.25em]
        
        Human accountability and control &
        Who configures, approves, supervises, overrides, or interprets the AI's output? &
        Staff approval gates; TA vetting; instructor-curated scenario banks; manual edits before publication; dashboards interpreted by course leaders &
        Edit rates; approval latency; staff trust; quality of final outputs; evidence that humans can intervene meaningfully \\
        
        \addlinespace[0.25em]
        
        \begin{tabular}[t]{@{}l@{}} Learner agency and\\cognitive engagement \end{tabular} &
        What intellectual work remains with the learner? &
        Reflection prompts; debugging questions; critique of AI output; scaffolded explanations; partial feedback; revision opportunities &
        Reflection quality; self-explanation; debugging behavior; transfer to later tasks; student perceptions of ownership and effort \\
        
        \addlinespace[0.25em]
        
        \begin{tabular}[t]{@{}l@{}} Context specificity\\and boundary\\setting \end{tabular} &
        How is the system localized to the course, assignment, population, and instructional norms? &
        Assignment context; course policies; local terminology; code-style conventions; instructor examples; scenario banks; topic restrictions &
        Course fit; reduced ambiguity; appropriateness across student groups; maintainability as course materials change \\
        
        \addlinespace[0.25em]
        
        Evaluation visibility and revision &
        How can instructors observe the system's behavior and improve the governance design over time? &
        Logs; dashboards; staff edits; student ratings; TA feedback; domain-level analytics; prompt and content versioning &
        Actionable analytics; identifiable failure modes; revision of prompts, rubrics, scenarios, or workflows; sustained adoption \\
        
        \bottomrule
    \end{tabularx}
    \caption{Dimensions of an instructional governance profile for genAI teaching tools.}
    \label{tab:profile-dimensions}
\end{table*}

The dimensions are grounded in a familiar premise from learning sciences: instructional tools should be evaluated by the learning activity they organize, not merely by the content they produce. Pedagogical grounding connects AI outputs to course goals and theories of learning, such as constructive alignment, self-regulated learning, metacognition, and technological pedagogical content knowledge \cite{biggs1996constructive,zimmerman2002becoming,krathwohl2002revision,mishra2006tpck}. AI instructional authority specifies how much pedagogical work the model is allowed to perform. Human accountability and control determine where course staff retain responsibility for consequential decisions. Learner agency and cognitive engagement specify the student's role in the interaction. Context specificity encodes local curricular norms into the system. Evaluation visibility turns governance from a one-time design decision into an inspectable improvement process.

We use the term \emph{governance fit} to describe coherence between a tool's instructional function and its governance profile. A forum-response assistant, for example, produces text that may become part of the official voice of the course; its governance fit therefore depends heavily on staff approval, edit traces, and accountability for publication. A reflection assistant, by contrast, may be most appropriately governed by pedagogical grounding and learner agency: the system's value lies in the quality of the thinking it elicits rather than in the authority of the answers it provides. A simulated-student environment for GTA training requires a different profile again: the AI may be authorized to perform as a student persona and provide rubric-aligned formative feedback, while program leaders retain responsibility for interpreting analytics and coaching instructors.

The framework can be used at three moments in the design cycle. First, before implementation, it prompts designers to specify the instructional function of the tool and the governance mechanisms needed for that function. Second, during deployment, it helps instructors monitor whether the tool is preserving the intended distribution of work among students, staff, and AI. Third, after use, it guides evaluation by identifying the evidence that matters for a particular profile. A tool governed through a rubric should be evaluated partly through rubric alignment and interpretability; a tool governed through an approval gate should be evaluated partly through edit traces, staff workload, and final response quality; a tool governed through learner reflection should be evaluated partly through the depth, detail, and usefulness of reflection it supports.

This profile-based view also provides a common vocabulary for comparing AI teaching tools that would otherwise be grouped together under broad labels such as ``AI tutor'' or ``AI teaching assistant.'' Such labels often hide the most important design choices. A tool that generates official course-forum responses, a tool that asks post-submission reflection questions, and a tool that simulates a frustrated student in an office-hour conversation may all use similar language models, but they distribute authority and accountability in very different ways. The governance profile makes those differences available for design, review, and research.

We constructed the framework through an abductive design-analysis process: first identifying recurring governance decisions across seven AI-mediated teaching tools developed in computing and first-year engineering contexts; then organizing those decisions using constructs from constructive alignment, self-regulated learning, TPACK, and human-AI interaction; and finally probing its transferability by applying it to four published AI-supported programming systems varying in dominant governance mechanism.

\section{Governance Profiles in a Portfolio of AI Teaching Tools} \label{sec:portfolio}

We illustrate the framework through a portfolio of AI teaching tools and supporting frameworks developed for computing and first-year engineering contexts. The cases span multiple forms of instructional work: preparing GTAs for student-facing conversations, supporting reflective programming practice, drafting course-forum responses, generating conceptual diagrams, coaching course-specific code style, onboarding students to responsible GenAI use, and assisting instructors with course-content development. This breadth is useful because it shows how governance profiles vary with instructional function rather than with the underlying AI model alone.

\Cref{tab:portfolio-profiles} summarizes the governance profiles of these cases. The cases should be read as design vignettes rather than as a single pooled intervention. Each case makes one governance mechanism especially visible: rubric anchoring, learning-theory grounding, staff approval, supervised review, course-specific constraints, responsible use onboarding, or reproducible instructor workflow.

\begin{table*}[t]
    \centering
    \scriptsize
    \begin{tabularx}{\textwidth}{p{0.07\textwidth}p{0.14\textwidth}p{0.20\textwidth}p{0.25\textwidth}p{0.25\textwidth}}
        \toprule
        \textbf{Case} &
        \textbf{Instructional function} &
        \textbf{Pedagogical grounding} &
        \textbf{AI authority and human control} &
        \textbf{Agency, context, and visibility} \\
        \midrule
        
        \glow &
        GTA rehearsal for office-hour and help-room interactions &
        Office-hour conversation rubric; instructor-authored scenario and persona design; GTA training goals &
        AI simulates student personas and provides rubric-aligned formative feedback; program staff interpret analytics and connect results to human coaching &
        GTAs practice in repeatable, low-risk scenarios; dashboards support domain-level coaching and programmatic revision \cite{dickey2026glow} \\
        
        \addlinespace[0.25em]
        
        \owlgorithm &
        Post-submission reflection in competitive programming &
        Self-regulated learning and Bloom-aligned reflection prompts; submission outcome and code context &
        AI generates metacognitive questions tailored to accepted, partial, or unsuccessful submissions; students remain responsible for reasoning and debugging &
        Learner agency is preserved through reflection rather than answer delivery; student ratings and TA feedback expose prompt quality and usability \cite{nieto2026owlgorithm} \\
        
        \addlinespace[0.25em]
        
        \boilertai &
        Instructional staff support for course discussion forums &
        Course norms for forum response quality, tone, and conceptual explanation &
        AI drafts candidate responses; staff review, edit, reprompt, and approve before publication &
        The official course voice remains staff-accountable; response time, staff edits, student satisfaction, and TA workload provide evaluation traces \cite{sinha2024boilertai} \\
        
        \addlinespace[0.25em]
        
        \boilersketch &
        Diagrammatic and multimodal support for conceptual CS topics &
        Instructor- and TA-vetted representations of common CS concepts; graph and diagram conventions &
        AI generates Mermaid-style diagrams and explanatory support; TAs supervise and vet outputs before instructional use &
        Governance is enacted through representational review; staff judgments make diagram quality visible for routine topics \cite{dickey2026boilersketch} \\
        
        \addlinespace[0.25em]
        
        \codestylist &
        Course specific code-style coaching for early CS students &
        Local style guides, course conventions, and expectations for readable code &
        AI suggests style revisions and explanations; instructors configure the norms against which suggestions are generated &
        Students retain ownership of code revision; evaluation can focus on style consistency, uptake, clarity, and calibration to course expectations \cite{dickey2026codestylist} \\
        
        \addlinespace[0.25em]
        
        \ailab &
        Structured onboarding for student use of GenAI in programming courses &
        Guided activities around prompting, critique, debugging, tool limitations, and mindful use &
        AI is both a tool and an object of critique; instructors frame appropriate uses and connect activities to course goals &
        Students practice evaluating AI output while reflecting on skill development; surveys and focus groups reveal shifts in perceptions and usage patterns \cite{dickey2024ailab,dickey2026ailabeval} \\
        
        \addlinespace[0.25em]
        
        \gaide &
        Instructor-facing course content development &
        Constructivist and TPACK-informed workflow for developing instructional materials &
        AI assists with brainstorming, drafting, variation, and refinement; instructors retain judgment over pedagogical quality and adoption &
        Governance is embedded in a reproducible prompt/content workflow; instructors can revise prompts, examples, and materials while preserving course intent \cite{dickey2024gaide} \\
        
        \bottomrule
    \end{tabularx}
    \caption{Governance profiles across a portfolio of AI teaching tools and supporting frameworks.}
    \label{tab:portfolio-profiles}
\end{table*}

\vspace{-1em}
\subsection{Governance Through Instructional Artifacts}
The first pattern across the portfolio is that governance is often carried by instructional artifacts that predate the AI system: rubrics, learning objectives, scenario banks, style guides, or content development workflows. These artifacts encode local values into the tool. They also provide a stable reference point for evaluating outputs that would otherwise be judged only by surface plausibility.

\glow illustrates rubric-anchored governance. The instructional problem is preparing GTAs for office-hour interactions that are simultaneously conceptual, procedural, and affective. The AI system is authorized to simulate student personas and provide formative feedback, constrained by an office-hour conversation rubric and interpreted through programmatic coaching. The rubric acts as a governance mechanism that defines good instructional interaction, makes performance domains visible, and provides a shared language for follow-up support. In this profile, AI-generated simulation is appropriate because the learning activity is rehearsal. Repeatability, standardization, and controlled variation are pedagogical assets.

\owlgorithm illustrates theory-grounded governance. Competitive programming environments often emphasize fast problem solving and objective correctness, which can leave limited space for reflection after a submission. \owlgorithm repositions AI from solution generator to reflection generator by using self-regulated learning and Bloom-aligned prompting. The system's authority is therefore intentionally focused: it produces metacognitive prompts that invite students to explain, debug, compare strategies, or articulate why a solution worked. The learner remains responsible for reasoning through the code. The governing artifact is the reflection schema, which shapes the AI's contribution toward self-explanation and post-task learning.

\codestylist illustrates course-standard governance. Code style is pedagogically important in early programming courses because it represents conventions of readability, maintainability, and professional practice; it is also highly local. A generic model can comment on style, but a course-specific style coach must draw its authority from the course's explicit standards. The governance profile therefore rests on context specificity: suggestions should reflect the course style guide, the assignment level, and the expectations instructors want novices to internalize. In this profile, AI acts as a formative coach. Its contribution is valuable when suggestions help students see and revise patterns in their own code while keeping the act of revision under the student's control.

\gaide illustrates workflow governance for instructors. Its central instructional function is not to teach students directly, but to help instructors create, adapt, and refine course materials. The relevant governance question is therefore how AI-assisted content development remains aligned with pedagogical intent. \gaide addresses this by embedding AI use in a reproducible workflow that connects prompts, course goals, content knowledge, and instructor judgment. The workflow makes the instructor's pedagogical reasoning part of the system, rather than treating content generation as an isolated model output.

\subsection{Governance Through Workflow Placement}
A second pattern is that governance can be enacted by placing the AI system at a specific point in the instructional workflow. \boilertai is the clearest example. Discussion forums are consequential because students often interpret staff responses as official course guidance. In this setting, the AI system is given authority to draft, vary, and help refine possible responses, while staff retain authority to approve what becomes visible to students. The approval gate is not an optional layer after AI generation; it is the core workflow. This placement lets the tool address a real scaling problem -- the time and cognitive load required to respond well to many student posts -- while preserving human accountability for the final instructional message.

\boilersketch uses a related but distinct form of workflow governance. The instructional function is diagrammatic explanation: translating handwritten or textual prompts into graphs, Mermaid diagrams, or conceptual visualizations that can support CS learning. Diagrams have a particular governance challenge because they can appear authoritative even when they encode a subtle misconception. The tool's governance profile therefore places TA supervision around diagram generation. The AI system can accelerate representational work, but staff remain responsible for judging whether a visualization is conceptually correct and pedagogically helpful. In this profile, supervision is especially appropriate for routine topics where staff can rapidly inspect the generated representation and decide whether it is suitable for student use.

Together, \boilertai and \boilersketch show that ``human in the loop'' is most useful when specified as a concrete workflow relation. Humans may configure the tool, vet outputs, edit drafts, approve publication, interpret analytics, or coach users after the interaction. These are different forms of control. The governance profile clarifies which form is needed for a given instructional function.

\subsection{Governance Through Learner Role Design}
A third pattern is that governance operates through the learner role the tool constructs. \ailab, \owlgorithm, and \codestylist all preserve meaningful student work by shaping AI use around critique, reflection, debugging, and revision.

\ailab treats GenAI as a tool students must learn to use deliberately. The learning goal is not simply exposure to AI, but the development of judgment about when AI output is useful, when it is misleading, and how it relates to core programming skill development. Students engage with prompts, inspect limitations, compare outputs, and reflect on responsible use. The governance profile places substantial emphasis on learner agency and cognitive engagement: students are expected to interrogate AI output rather than merely consume it. The instructor's role is to design the activity sequence and connect the experience to course norms around learning, practice, and independence.

\owlgorithm designs the learner role around post-submission reflection. A correct submission can prompt explanation of strategy and complexity; a partial or failed submission can prompt debugging, comparison, and planning. In each case, the AI system creates an opportunity for reflection while leaving the central learning work with the student. \codestylist similarly positions AI feedback as an invitation to revise. The system identifies style issues and explains course-specific expectations, but the learner must decide how to improve the program. These tools demonstrate learner agency is not abstract; it is an interaction pattern that can be designed.

\subsection{Governance Through Visibility and Revision}
The final cross-case pattern is that governance becomes stronger when it is visible to instructors and revisable. \glow's rubric-domain analytics, \boilertai's edit and approval traces, \owlgorithm's student ratings and TA feedback, \ailab's survey and focus-group evidence, and \boilersketch's staff review all make aspects of tool behavior inspectable. These traces allow instructors to revise rubrics, prompts, scenarios, diagrams, approval practices, and onboarding.

This visibility also changes what counts as evidence. A single portfolio cannot be evaluated through one common outcome measure because each tool performs a different instructional function. Instead, each governance profile points to the evidence that matters. For \glow, relevant evidence includes rubric-domain growth and GTA readiness for complex student interactions. For \owlgorithm, relevant evidence includes reflection quality, debugging usefulness, and prompt alignment. For \boilertai, relevant evidence includes response timeliness, final answer quality, staff edit effort, and student satisfaction. For \boilersketch, relevant evidence includes representational clarity and staff confidence after review. For \ailab, relevant evidence includes shifts in students' comfort, mindfulness, and ability to articulate appropriate GenAI use. The framework therefore links design choices to evaluation choices.

Across the portfolio, the central lesson is that responsible scale is achieved by matching the governance profile to the instructional work. Rubric-anchored simulation supports GTA rehearsal; learning-theory-grounded prompting supports reflection; approval gates support official course communication; supervised review supports diagrammatic explanation; course-specific standards support style coaching; structured onboarding supports mindful student use; and reproducible workflows support instructor content development. Rather than viewing these as competing levels of governance, we study them as different configurations of authority, agency, accountability, context, and visibility.

\section{Applicability to Other genAI Teaching Tools} \label{sec:external-tools}
A useful framework should classify systems beyond the cases that motivated it. We therefore apply the governance-profile lens to four published AI-supported programming and computational-learning tools selected for contrast: CodeHelp \cite{liffiton2024codehelp}, CodeAid \cite{kazemitabaar2024codeaid}, CodeTailor \cite{hou2024codetailor}, and BISCUIT \cite{cheng2024biscuit}. Results can be seen in \Cref{tab:external-profiles}. These cases are not intended as an exhaustive taxonomy of AI tools in computing education. They function as a transfer test: each tool uses genAI to support programming or computational work, but each governs the interaction through a different mechanism.

\begin{table*}[t]
    \centering
    \scriptsize
    \renewcommand{\arraystretch}{1.12}
    \begin{tabularx}{\textwidth}{@{}p{0.05\textwidth}p{0.13\textwidth}p{0.13\textwidth}X p{0.18\textwidth}@{}}
        \toprule
        \textbf{Case} &
        \textbf{Instructional function} &
        \textbf{Dominant governance mechanism} &
        \textbf{Governance-profile classification} &
        \textbf{Transferable framework insight} \\
        \midrule
        
        CodeHelp &
        On-demand programming help for students while preserving problem-solving effort &
        \textbf{Guarded assistance}: governing the boundary between help and answer delivery &
        \textbf{Pedagogical grounding:} programming-help norms that distinguish productive guidance from direct solution disclosure.
        \textbf{AI authority/control:} the AI may explain, hint, and guide within constrained response boundaries; instructors decide whether those boundaries fit the course context.
        \textbf{Agency/context/visibility:} students remain responsible for constructing and debugging their own solutions; evaluation should examine helpfulness, solution leakage, reliance, and educator trust. &
        Guardrails are not only safety features, they are learner-agency mechanisms. The reusable question is: \emph{what help boundary preserves the intended student work?} \cite{liffiton2024codehelp} \\
        
        \addlinespace[0.25em]
        
        CodeAid &
        Classroom deployed programming assistant balancing student support and educator concerns &
        \textbf{Balanced classroom assistance}: governing both outputs and interaction patterns at course scale &
        \textbf{Pedagogical grounding:} large-course support needs are balanced against concerns about over-assistance, correctness, transparency, and student control.
        \textbf{AI authority/control:} the AI provides programming assistance through a deployed interface; human control is expressed through system constraints, transparency choices, and course-level expectations.
        \textbf{Agency/context/visibility:} governance must address what students can request, what the system can reveal, and how the interaction remains legible to educators. &
        At scale, governance cannot focus only on generated content. It must also govern the request space, the response space, and the visibility instructors have into the interaction \cite{kazemitabaar2024codeaid}. \\
        
        \addlinespace[0.25em]
        
        CodeTailor &
        Personalized Parsons puzzle generation for programming support &
        \textbf{Task transformation}: governing assistance by changing its instructional form &
        \textbf{Pedagogical grounding:} Parsons problems provide a scaffolded form of code construction, ordering, and reasoning rather than direct answer receipt.
        \textbf{AI authority/control:} the AI transforms student context into a structured practice artifact; human control is primarily curricular, through the decision to route assistance into this task form.
        \textbf{Agency/context/visibility:} students must assemble, reason about, and test a solution; the generated puzzle provides an inspectable artifact for evaluating whether the scaffold is appropriate. &
        The framework should classify not only how much information the AI gives, but what kind of learning activity the AI creates. A reusable question is: \emph{can answer-seeking be transformed into scaffolded construction?} \cite{hou2024codetailor} \\
        
        \addlinespace[0.25em]
        
        BISCUIT &
        LLM-supported code generation in computational notebooks &
        \textbf{Interface mediation}: governing AI authority through staged user control &
        \textbf{Pedagogical grounding:} computational-notebook work requires coordination among natural-language intent, generated code, execution, and interpretation.
        \textbf{AI authority/control:} the AI can generate code, but an ephemeral interface mediates between user intent and accepted code; users inspect, steer, and structure generation before incorporation.
        \textbf{Agency/context/visibility:} the intermediate interface creates a visible space for interpretation before generated code becomes part of the notebook. &
        Interfaces can be governance mechanisms. The reusable question is: \emph{where should the tool slow, stage, or mediate the path from intent to executable artifact?} \cite{cheng2024biscuit} \\
        
        \bottomrule
    \end{tabularx}
    \caption{Applying the governance-profile framework to selected external AI teaching tools. Each row identifies the instructional function, names the dominant governance mechanism, maps that mechanism onto governance-profile dimensions, and extracts a transferable design question.}
    \label{tab:external-profiles}
\end{table*}

CodeHelp and CodeAid are useful comparison cases because they foreground guardrails in programming help. In both systems, the design challenge is to provide timely, scalable support while preserving student engagement with the programming task. From the governance-profile perspective, the key mechanism is the boundary placed around the AI's instructional authority, rather than merely the presence of an LLM in this context. The system may explain concepts, annotate mistakes, suggest strategies, or provide pseudocode-like guidance, while constraining responses that would substitute for the student's own solution construction. This boundary is a governance decision because it defines the learner's remaining work. It also suggests corresponding evidence: response helpfulness, direct solution leakage, student perceptions of control, student reliance patterns, and educator trust.

CodeTailor governs the interaction differently. Rather than constraining a natural-language response alone, it transforms assistance into a personalized Parsons puzzle. This design changes the artifact through which help is given. The AI system uses student context to generate a structured practice activity, and the learner must still assemble, reason about, and test a solution. In the language of our framework, CodeTailor's strongest governance dimension is learner agency through task design. The tool's contribution is that it redirects the interaction toward an established pedagogical form that requires active construction, as opposed to merely withholding the answer. Thus, governance profiles should attend to representational form, not only content filtering or approval.

BISCUIT provides a third form of governance: interface mediation. In computational notebooks, LLM-generated code can move quickly from natural language to executable artifact. BISCUIT inserts an ephemeral UI between the user's intention and the generated code, allowing the user to inspect, guide, and structure the generation process. This is a governance profile in which human control is mediated by interface design. The AI system retains the capacity to generate code, but its authority is staged through an intermediate artifact that invites user interpretation before code is accepted. The governance mechanism is therefore neither a rubric nor a staff approval gate; it is an interaction layer that slows down and structures the movement from intent to executable output.

These external examples demonstrate the value of classifying AI teaching tools by governance profile rather than by generic tool category. CodeHelp, CodeAid, CodeTailor, and BISCUIT could all be described as AI-supported programming tools, but their governance mechanisms differ substantially. CodeHelp and CodeAid emphasize guarded assistance; CodeTailor emphasizes scaffolded task transformation; BISCUIT emphasizes interface-mediated control. The same framework also distinguishes these tools from the portfolio cases in \Cref{sec:portfolio}: \boilertai governs official communication through staff approval, \glow governs GTA rehearsal through rubric-anchored simulation, \owlgorithm governs reflection through metacognitive prompting, and \gaide governs instructor content development through reproducible workflow.

The classification also reveals how future systems can be designed more deliberately. A programming tutor for novices might combine CodeAid-style guardrails with CodeTailor-style task transformation. A staff-facing tool might combine \boilertai-style approval gates with \gaide-style workflow documentation. A code-generation environment might combine BISCUIT-style interface mediation with course-specific constraints similar to \codestylist. The governance-profile framework supports such design reasoning by making the mechanisms composable: rubrics, prompts, approval gates, scaffolds, style guides, interfaces, and review processes can be selected and combined according to the instructional function.

The broader implication is that responsible AI integration in computing education should be studied at the level of governance mechanisms. Questions such as whether students ``use AI'' or whether a tool is an ``AI tutor'' are too coarse to guide design or evaluation. More productive questions include: What authority does the AI system have in this interaction? What intellectual work does the learner retain? Which human role is accountable for final outputs? What course-specific artifacts constrain the system? What traces allow the tool to be improved? These questions move the field toward testable design claims about AI-mediated learning environments and provide a shared vocabulary for educators, researchers, and tool builders.


\bibliographystyle{ACM-Reference-Format}
\bibliography{refs}

\end{document}